\documentclass[a4paper,11pt,fleqn,epsfig]{article}

\usepackage[affil-it]{authblk}
\usepackage{latexsym}
\usepackage{epstopdf}
\usepackage{graphicx}
\usepackage{amssymb}
\usepackage{amsmath}
\usepackage{amsfonts}

\begin{document}

\title{Raising and lowering operators for quantum billiards}

\author{Ayush Kumar Mandwal}
\affil{\small Complexity Science Group, University of Calgary, Alberta T2N 1N4, Canada}

\author{Sudhir R. Jain%
\thanks{Electronic address: \texttt{srjain@barc.gov.in}; Tel.: +912225593589}}
\affil{\small Nuclear Physics Division, Bhabha Atomic Research Centre, Mumbai 400085, India}
\affil{\small Homi Bhabha National Institute, Anushakti Nagar, Mumbai 400094, India}

\date{Dated: \today}

\maketitle

\begin{abstract}
For planar integrable billiards, the eigenstates can be classified with respect to a quantity determined by the quantum numbers. Given the quantum numbers as $m, n$, the index which represents a class is $c = m\, \mbox{mod}\, k\,n$ for a natural number, $k$. We show here that the entire tower of states can be generated from an initially given state by application of the operators introduced here. Thus, these operators play the same role for billiards as raising and lowering operators in angular momentum algebra. 

\end{abstract}

Quantum billiards are systems where a single particle is confined inside a boundary on which the eigenfunctions vanish \cite{gutzwiller}. One seeks the solutions of the  time-independent Schr\"{o}dinger equation, which is the same as the Helmholtz equation in the context of general wave phenomena. The solutions of this problem for an arbirarily shaped enclosure is a very challenging open problem, even when we restrict ourselves to two-dimensional cases \cite{jgk,jl,j2008,j2009}. There are some very interesting connections between exactly solvable models and random matrix theories, a summary may be seen in \cite{czech}. The Helmholtz operator is separable in certain coordinate systems - for these cases, the solutions can be found \cite{manjunath}. The non-separable problems for which the classical dynamics is integrable have been recently studied in detail \cite{amit,rhine1,rhine2}. Although the solutions of these systems have been known, there remain many questions regarding the nature of nodal curves and domains. The nodal domains of the eigenfunctions of the systems for which the Schr\"{o}dinger equation is separable, form a checkerboard pattern \cite{manjunath}. the number of crossings actually count the number of domains. Moreover, the checkerboard patterns are trivially self-similar.
  
\begin{figure}
\centering
\begin{minipage}{.3\textwidth}
  \centering
  \includegraphics[scale=0.3,height=33mm]{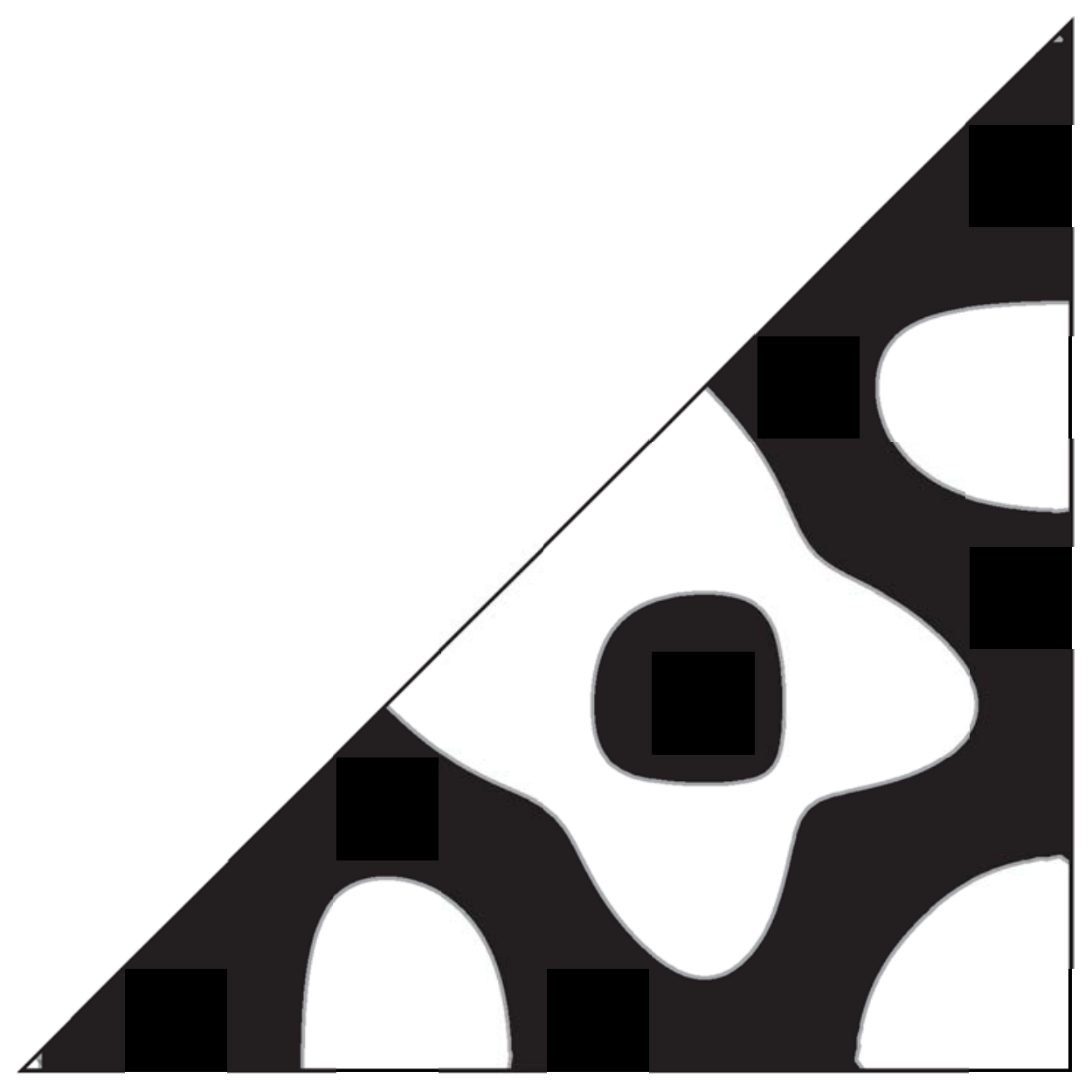}
  (a)
\end{minipage}
\begin{minipage}{.3\textwidth}
  \centering
  \includegraphics[scale=0.3,height=33mm]{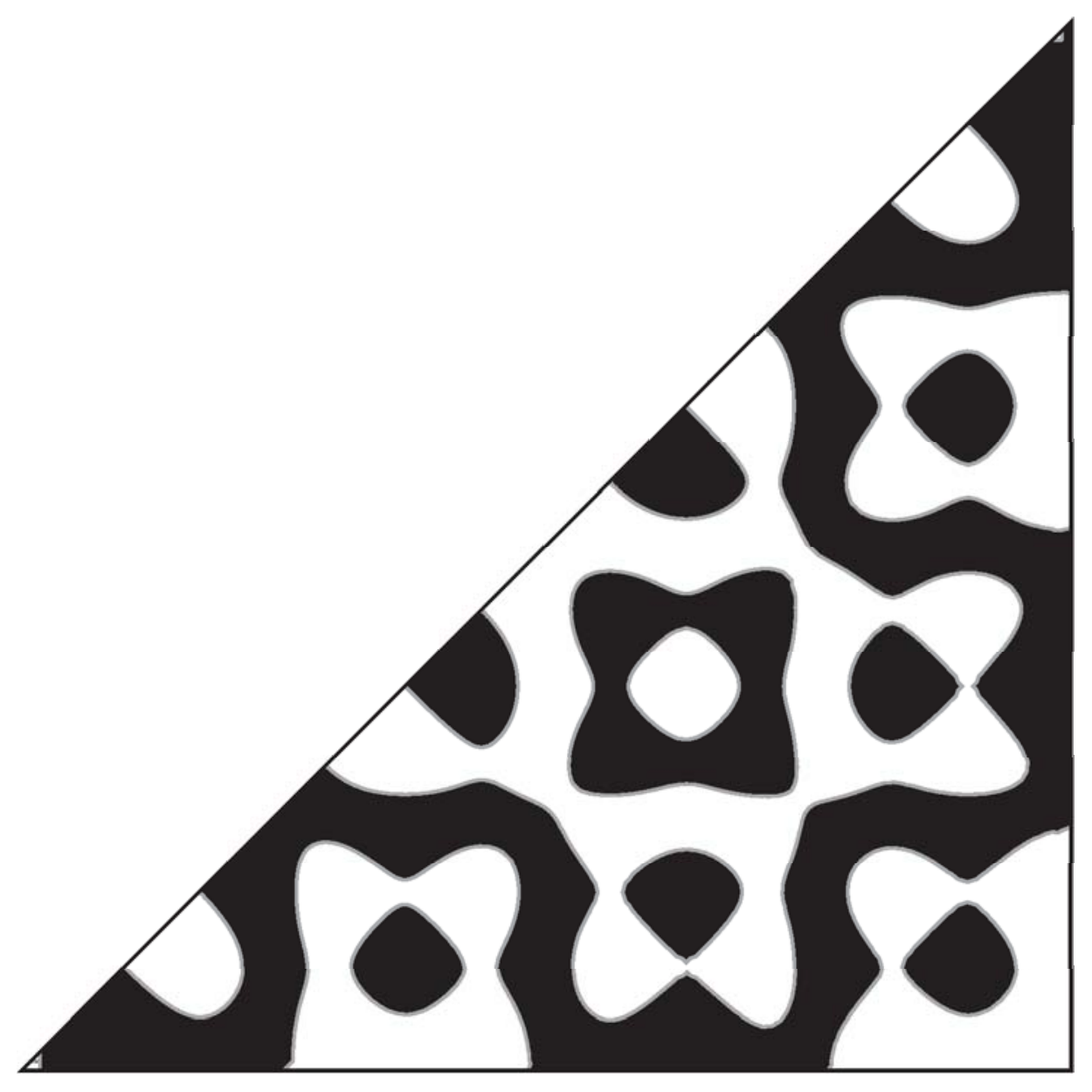}
  (b)
\end{minipage}
\begin{minipage}{.3\textwidth}
  \centering
  \includegraphics[scale=0.3,height=33mm]{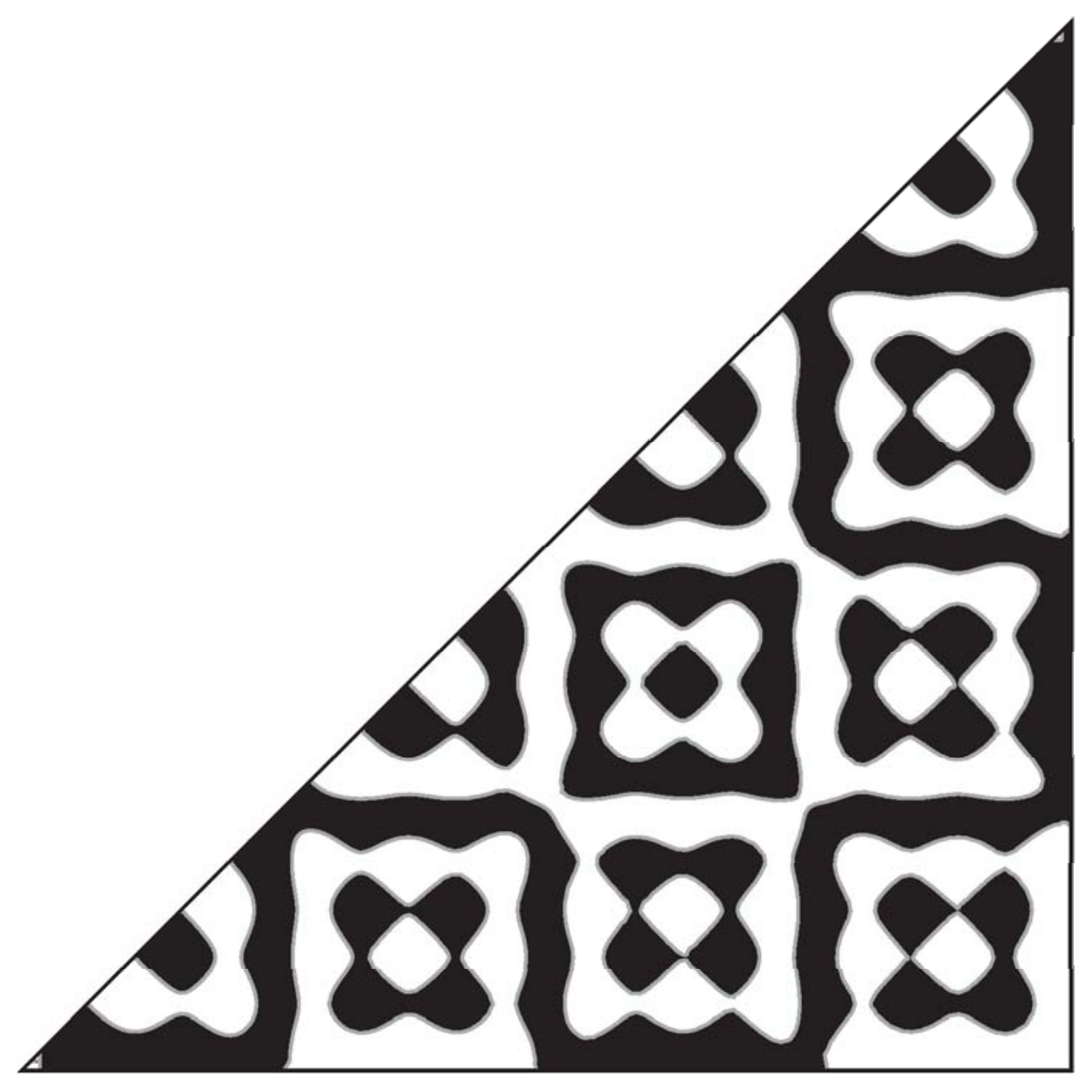}
  (c)
\end{minipage}

\caption{The pattern of nodal domains of the right-angled isosceles triangle for (a) $\psi_{\,7,4}$, (b)  $\psi_{\,15,4}$ and (c)  $\psi_{\,23,4}$. All three eigenfunctions belong to the same equivalence class $\big[{\cal C}_{2n}\big]$ and the similarity of the nodal pattern is evident as the wavefunction evolves from one state to another within members of the same class.}
\label{fig:iso}
\end{figure}

Counting the nodal domains of non-separable plane polygonal billiards is very difficult in general \cite{berry}. Even if we restrict to systems that are classically integrable, the problem poses considerable challenge. Progress on this otherwise intractable problem could be made recently due to the observation that the eigenfunctions could be classified in terms of equivalence classes \cite{rhine1,manjunath}. Fig. \ref{fig:iso} shows examples of eigenfunctions belonging to an equivalence class  in the right isosceles triangle billiard. One cannot miss the remarkable similarity in each family, they seem genetically related. Here we shall present operators that make any other state appear starting from one in a family. Thus, we can construct the tower of states by repeated application of this operator. This reminds us of the usual raising and lowering operators in quantum mechanics. 

We explain in the following Sections the construction of ``raising" and ``lowering" operators for the right isosceles and equilateral triangle billiard, and summarize with remarks about other systems.

\section{Right Isosceles Triangle}  

The solutions of the Helmholtz equation for the right isosceles triangle with sidelength, $\pi$ (chosen for convenience) are given by 
\begin{equation}\label{eq:iso}
\psi _{m,n}(x,y)=\sin (mx)\sin (ny)-\sin (nx)\sin (my),
\end{equation}
$m > n$. This consists of two terms, each being a product of $\sin $ functions. Of course, it can be re-written in a variety of equivalent ways by employing trigonometric identities. With just one term of a product of sine functions, the nodal lines are straight lines and they form a checkerboard pattern. This would be the case also for a product of any other special  function.\\

For instance, the solutions of the Helmholtz equation for a circular, elliptical, circular annulus, elliptical annulus, confocal parabolic enclosures are each a  product of functions like Bessel for circular, Mathieu for elliptic and so on \cite{manjunath}. \\

Eq. (\ref{eq:iso}) can be rewritten in a way that will be more useful:  

\begin{eqnarray}\label{eq:iso1}
\psi _{m,n}(x,y) &=&\frac{1}{2}\Re \left[ e^{i(mx-ny)}-e^{i(mx+ny)}-e^{i(my-nx)}+e^{i(my+nx)} \right] \nonumber \\
&=& \frac{1}{2} \Re {\rm Tr~}
\left[\begin{array}{cc}
\{e^{i(mx-ny)}-e^{i(mx+ny)}\} & 0\\
0 & \{-e^{i(my-nx)}+e^{i(my+nx)}\}
\end{array}\right] \nonumber \\ &:=& \frac{1}{2} \Re {\rm Tr~} {\mathcal I}. 
\end{eqnarray}

All the eigenfunctions can be classified into equivalence classes labelled by ${\cal C}_{2n} = m {\rm mod~} 2n$ \cite{rhine1}. Within each class, it was shown that the number of domains, $\nu _{m,n}$ for one eigenfunction is related to $\nu _{m+2n,n}$ by a difference equation \cite{rhine1,manjunath}. We can, in fact, write down the operator (in the matrix form) which actually takes us along the ladder of states beginning with $\psi _{m,n}$, up and down. The matrix is 
\begin{equation}\label{eq:op_iso}
{\mathcal T} = \left[\begin{array}{cc}
e^{i2nx} & 0\\
0 & e^{i2ny}
\end{array}\right].
\end{equation}

To confirm, we get the eigenfunction $\psi_{m+2np,n}(x,y)$ as
\begin{equation}\label{eq:isosceles}
\psi_{m+2np,n}(x,y)=\frac{1}{2} \Re {\rm Tr~} \left( {\mathcal T}^p {\mathcal I} \right).
\end{equation}
Thus, we have generated all the states beginning anywhere; note that $p$ could be any integer as long as we keep the inequality between the two quantum numbers.

\section{Equilateral triangle}

The eigenfunctions of an equilateral triangle of side length $\pi$, satisfying the Dirichlet boundary conditions, can be written as three terms, each a product of trigonometric functions \cite{rhine2}. There are two possible solutions - one with cosine and th other with sine functions. First we discuss the function with cosines:  
\begin{eqnarray}\label{eq:equi}
\psi _{m,n}^{c}(x,y) &=& \cos((2m-n)\frac{2x}{3})\sin(n\frac{2y}{\sqrt{3}})-\cos((2n-m)\frac{2x}{3})\sin(\frac{2m}{\sqrt{3}}y) \nonumber \\ &+& \cos((m+n)\frac{2x}{3})\sin((m-n)\frac{2y}{\sqrt{3}}).
\end{eqnarray}
This can be re-written as
\begin{eqnarray}\label{eq:equi1}
\psi _{m,n}^{c}(x,y) &=& \frac{1}{2}\Im \bigg[ e^{i\frac{4}{3}(mx-n(\frac{x}{2}-\frac{\sqrt{3}y}{2}))}-e^{i\frac{4}{3}(mx-n(\frac{x}{2}+\frac{\sqrt{3}y}{2}))} \nonumber \\ &-& e^{i\frac{4}{3}(nx-m(\frac{x}{2}-\frac{\sqrt{3}y}{2}))} + e^{i\frac{4}{3}(nx-m(\frac{x}{2}+\frac{\sqrt{3}y}{2}))}
\nonumber \\
&-& e^{i\frac{4}{3}(m(\frac{x}{2}-\frac{\sqrt{3}}{2}y)+n(\frac{x}{2}+\frac{\sqrt{3}y}{2}))}+e^{i\frac{4}{3}(m(\frac{x}{2}+\frac{\sqrt{3}}{2}y)+n(\frac{x}{2}-\frac{\sqrt{3}y}{2}))} \bigg] \nonumber \\
&=& \Im \frac{1}{2}{\rm Tr~}{\mathcal A}
\end{eqnarray}
where ${\mathcal A}$ is 
\begin{eqnarray}
{\mathcal A} =\left[\begin{array}{ccccc}
\frac{e^{i\frac{4}{3}mx}}{e^{i\frac{4}{3}n(\frac{x}{2}-\frac{\sqrt{3}y}{2})}}-\frac{e^{i\frac{4}{3}mx}}{e^{i\frac{4}{3}n(\frac{x}{2}+\frac{\sqrt{3}y}{2})}}\\
 & \frac{-e^{i\frac{4}{3}nx}}{e^{i\frac{4}{3}m(\frac{x}{2}-\frac{\sqrt{3}y}{2})}} & & \text{\Huge 0}\\
 &  & \frac{e^{i\frac{4}{3}nx}}{e^{i\frac{4}{3}m(\frac{x}{2}+\frac{\sqrt{3}y}{2})}}\\
 \text{\Huge 0}&  &  & -\frac{e^{i\frac{4}{3}m(\frac{x}{2}-\frac{\sqrt{3}}{2}y)}}{e^{-i\frac{4}{3}n(\frac{x}{2}+\frac{\sqrt{3}y}{2})}}\\
 &  &  &  & \frac{e^{i\frac{4}{3}m(\frac{x}{2}+\frac{\sqrt{3}}{2}y)}}{e^{-i\frac{4}{3}n(\frac{x}{2}-\frac{\sqrt{3}y}{2})}}
\end{array}\right]
\end{eqnarray}
The matrix operator for this state is 
\[
T^{p}=\left[\begin{array}{ccccc}
e^{i4npx}\\
 & e^{-i4np(\frac{x}{2}-\frac{\sqrt{3}y}{2})} & & \text{\Huge 0}\\
 &  & e^{-i4np(\frac{x}{2}+\frac{\sqrt{3}y}{2})}\\
 &  \text{\Huge 0}&  & e^{i4np(\frac{x}{2}-\frac{\sqrt{3}y}{2})}\\
 &  &  &  & e^{i4np(\frac{x}{2}+\frac{\sqrt{3}y}{2})}
\end{array}\right]
\]
Similarly for the eigenfunctions written in terms of sine functions, 
\begin{eqnarray}
\Psi_{m,n}^{s}(x,y) &\sim & \sin \left( (2m-n)\frac{2x}{3} \right) \sin \left( n\frac{2y}{\sqrt{3}}\right) - \sin \left( (2n-m)\frac{2x}{3}\right) \sin \left( \frac{2m}{\sqrt{3}}y\right) \nonumber \\ &-&\sin \left( (m+n)\frac{2x}{3}\right) \sin \left( (m-n)\frac{2y}{\sqrt{3}} \right).
\end{eqnarray}
In complex form, it can be re-written as
\begin{eqnarray}
\Psi_{m,n}^{s}(x,y) &\sim & \frac{1}{2}\Re [e^{i\frac{4}{3}(mx-n(\frac{x}{2}+\frac{\sqrt{3}y}{2}))}-e^{i\frac{4}{3}(mx-n(\frac{x}{2}-\frac{\sqrt{3}y}{2}))}-e^{i\frac{4}{3}(nx-m(\frac{x}{2}+\frac{\sqrt{3}y}{2}))}+e^{i\frac{4}{3}(nx-m(\frac{x}{2}-\frac{\sqrt{3}y}{2}))} \nonumber \\
&-&e^{i\frac{4}{3}(m(\frac{x}{2}-\frac{\sqrt{3}}{2}y)+n(\frac{x}{2}+\frac{\sqrt{3}y}{2}))}+e^{i\frac{4}{3}(m(\frac{x}{2}+\frac{\sqrt{3}}{2}y)+n(\frac{x}{2}-\frac{\sqrt{3}y}{2}))}]
\end{eqnarray}	
and in matrix form as 
\begin{equation}
\Psi_{m,n}^{s}(x,y)=\frac{1}{2} \Re Tr [A].
\end{equation}
where $A$ is 
\[
A=\left[\begin{array}{ccccc}
-\frac{e^{i\frac{4}{3}mx}}{e^{i\frac{4}{3}n(\frac{x}{2}-\frac{\sqrt{3}y}{2})}}+\frac{e^{i\frac{4}{3}mx}}{e^{i\frac{4}{3}n(\frac{x}{2}+\frac{\sqrt{3}y}{2})}}\\
 & \frac{e^{i\frac{4}{3}nx}}{e^{i\frac{4}{3}m(\frac{x}{2}-\frac{\sqrt{3}y}{2})}} & & \text{\Huge 0}\\
 &  & -\frac{e^{i\frac{4}{3}nx}}{e^{i\frac{4}{3}m(\frac{x}{2}+\frac{\sqrt{3}y}{2})}}\\
 \text{\Huge 0}&  &  & -\frac{e^{i\frac{4}{3}m(\frac{x}{2}-\frac{\sqrt{3}}{2}y)}}{e^{-i\frac{4}{3}n(\frac{x}{2}+\frac{\sqrt{3}y}{2})}}\\
 &  &  &  & \frac{e^{i\frac{4}{3}m(\frac{x}{2}+\frac{\sqrt{3}}{2}y)}}{e^{-i\frac{4}{3}n(\frac{x}{2}-\frac{\sqrt{3}y}{2})}}
\end{array}\right]
\]
The corresponding matrix operator is
\[
T^{p}=\left[\begin{array}{ccccc}
e^{i4npx}\\
 & e^{-i4np(\frac{x}{2}-\frac{\sqrt{3}y}{2})} & & \text{\Huge 0}\\
 &  & e^{-i4np(\frac{x}{2}+\frac{\sqrt{3}y}{2})}\\
 &  \text{\Huge 0}&  & e^{i4np(\frac{x}{2}-\frac{\sqrt{3}y}{2})}\\
 &  &  &  & e^{i4np(\frac{x}{2}+\frac{\sqrt{3}y}{2})}
\end{array}\right]
\]
This operator is the same as for the cosine form of the eigenfunctions for equilateral triangle billiard.

\section{Conclusions}

The eigenfunctions of separable billiards are a single product of special functions - trigonometric for rectangular billiard, Bessel and trigonometric functions for circular billiards (and related annuli), Mathieu and trigonometric functions for elliptical billiards (and annuli), and parabolic cylinder functions for confocal parabolic billiards. In all these cases, the tower of states can be trivially constructed along the lines described here. This is because the index that classifies states for all separable billiards is (${\mathcal C}_{n} = m \mod n$). For the non-separable billiards described here, we have shown in earlier papers that all the states can be classified by (${\mathcal C}_{2n} = m \mod 2n$) or (${\mathcal C}_{3n} = m \mod 3n$). Here, we have shown that within a class, all the states can be constructed from the energetically lowest state. We can also make a transformation from an excited state to the lowest state. We hesitate to call this a `ground state' as there will be one lowest state for an index, ${\mathcal C}_{k\,n}$, $k = 1, 2, 3$.   

The results given here are for billiards with Dirichlet boundary conditions. Of course, these results are trivially extended to the case of periodic boundary conditions. The raising and lowering operators will remain the same. For twisted boundary conditions, these may be generalized by introducing phases in the matrix representation of raising and lowering operators.

\end{document}